\documentclass[amsmath]{article}
\usepackage{amssymb}
\usepackage{amsfonts}
\usepackage[fleqn]{amsmath}
\usepackage{ulem}
\usepackage{graphicx,epsfig}%
 \newtheorem{theorem}{Theorem}[section]
 
 \newtheorem{proposition}[theorem]{Proposition}

 \newtheorem{Definition}[theorem]{Definition}

 \newenvironment{proof}[1][Proof]{\begin{trivlist}
 \item[\hskip \labelsep {\bfseries #1}]}{\end{trivlist}}

\date{}
\begin{document}
\title{Partial thermalization in the quantum chain \\of harmonic oscillators}
\author{S.~B.~Rutkevich \footnote {
e-mail: sergei.rutkevich@uni-duisburg-essen.de
} 
\\
{\small{ 
SSPA "Scientific-Practical Materials Research Centre, 
NAS of Belarus", Minsk, Belarus
}}  \\
{\small{ and}}\\
{\small{Fakult\"at f\"ur Physik, Universit{\"a}t Duisburg-Essen,
D-47048 Duisburg, Germany
}}  }
\maketitle
\begin{abstract}
This preprint contains the English translation of the paper 
"Relaxation dynamics of a quantum chain of harmonic oscillators", which was 
published by the author in 1980 in the Ukrainian Physical Journal. Comments
describing its motivations, background ideas and results  
are presented as well. This paper addressed to the problem of 
approach to the thermal equilibrium in an isolated macroscopic quantum system,
which was studied on the example of  the
quantum chain of weakly interacting harmonic oscillators. In the initial state,
macroscopic energy was supplied to one oscillator (atom) in the chain. 
Subsequent evolution of the quantum state of the whole chain was determined due
to the model integrability. 
The main subject
of interest was the time evolution 
of the reduced density operators characterizing
the quantum state of a particular atom. On the short time-scale, the energy perturbation
expands along the chain with the velocity of the fastest phonon mode. On the 
long-time scale, the single-atom density operators display strong fluctuations 
around the canonical Gibbs distribution. These fluctuations are caused by the 
degeneracy of the energy level differences (presence of resonances) 
in the model of coupled harmonic oscillators. After lifting this degeneracy, fluctuations are
suppressed providing, that the reduced density operator of each atom in the chain
becomes very close to the Gibbs distribution at almost any time moment.  
\end{abstract}
\section*{Introduction}
The claim that an isolated (closed) macroscopic system evolves to the 
thermal equilibrium state lies in the very basis of the statistical mechanics.
The question about the precise meaning of  this statement 
for classical and quantum systems 
arises for every student studying the university course of 
statistical mechanics.  With me this happens in Kharkov University in 1978.
Two years later, I published two papers \cite{rut80-1,rut80-2} in Russian-language 
journals, in which relaxation of the isolated macroscopic quantum system 
to the thermal equilibrium was studied on the example of the chain of harmonic oscillators. 
It turns out, that the set of problems considered 
in these papers and underlying ideas became the subject of considerable 
theoretical interest in subsequent years, to much extent, due to 
experiments on quantum dynamics in the  ultracold atomic gases 
\cite{Gr02,Bloch08,Kin06}. 
This relates, in particular, to the quantum quench  problem 
\cite{CC06,CC07,Olsh08,Silva09,Caz09,Silva11,CE11}, and to the equilibration 
scenario in quantum systems based on the {\it Eigenstate Thermalization 
Hypothesis} (ETH) \cite{Deutsch91,Sred94} and on the idea of canonical typicality 
\cite{Po06,Lin09,Leb06,Leb10a,Leb10b,Tas98}. 

While the second paper \cite{rut80-2} was later translated into English, the first one 
\cite{rut80-1} remained without translation. 
The aim of the present retrospective publication
is  to present the English translation of paper \cite{rut80-1}, 
complemented with  comments relating to its motivations and results.

In the both papers \cite{rut80-1,rut80-2}, the isolated
 linear chain of $N\gg1$ weakly 
coupled quantum harmonic oscillators ('the atoms')  was considered. The model Hamiltonian reads
as
\begin{equation}\label{HC}
H=\sum_{k=0}^{N-1}\frac{1}{2}\left[p_k^2+\Omega^2\,(q_k-q_{k+1})^2+ \Omega_0^2\, q_k^2\right],
\quad \Omega\ll\Omega_0.
\end{equation}
Here index
$k=0,\ldots,N-1,$ $N$ enumerates the atoms in the chain, $q_k$ and $p_k$ denote the special coordinate and 
momentum 
of the $k$-th atom, periodical boundary conditions are implied. 
At $t=0$, the chain was taken in the 
inhomogeneous pure  state $ |\psi(0)\rangle $, in which the macroscopic energy 
$E= {\mathfrak N} \Omega_0$, with 
$\bar{n}={\mathfrak N}/N \sim  1$ was supplied to one atom with $k=0$.
At $t>0$, the chain evolved into the  state $|\psi(t)\rangle$, which  was determined due to exact 
integrability of the model (\ref{HC}). 

In the first paper \cite{rut80-1}, the time evolution of the reduced density operator 
$\rho(k,t)$ of the atom located at the $k$-th site of the chain was calculated and studied,   
\begin{equation}\label{rdk}
\rho(k,t) = { \rm Tr }^{(r)}  \, \rho(t),   \quad 
\end{equation}
where $k=0,\dots,N-1$,  $ \rho(t) =| \psi(t) \rangle\langle \psi(t)|$ is the projector-type density operator 
of the whole chain, and ${\rm Tr }^{(r)}$ denotes the trace over the space corresponding to the rest 
$N-1$ atoms. 

On the short-time scale, the slow spreading of the energy from the initially excited atom along the chain 
with the velocity $\Omega^2/\Omega_0$ was found. On the other hand, the time average of the reduced 
density operators $\rho(k,t)$ of each atom in the chain  over the infinite time interval was found to 
approach to the Gibbs distribution in the thermodynamic limit $N\to\infty$. It turns out, however, that the 
time-fluctuations of the single-atom reduced density operators $\rho(k,t)$ remain considerable 
on the long-time scale.
Such atypical behavior was associated with 
the  strong degeneracy of the energy level
differences $(E_n-E_{n'})$ in model (\ref{HC}). It was shown in \cite{rut80-1}, that lifting this degeneracy
by arbitrary weak nonlinear interaction provides suppression of these fluctuations and guaranties approach
of the single-atom density operators $\rho(k,t)$ to the Gibbs distribution at large $t$. Though this result was 
proved only for the case $\bar{n}<1$, it is very likely, that it  holds also for $\bar{n}\ge 1$. 

It was shown in \cite{rut80-1} as well, that for a huge majority of eigenstates $|n\rangle$ of the 
total Hamiltonian
(\ref{HC}), the reduced density operators $\rho(k,n)={ \rm Tr }^{(r)}  
\, |n\rangle\langle n|$ corresponding to each individual $k$-th atom in the chain are equal to
the Gibbs distribution in the thermodynamic limit. In other words, a typical exact pure
stationary state $|n\rangle$  of the chain Hamiltonian would be perceived as the thermal equilibrium
state by the observer, which can see just one atom in the chain. This result agrees with the ETH, 
which was put forward later by Deutsch \cite{Deutsch91} and Srednicki \cite{Sred94}.

In paper \cite{rut80-2} the results of \cite{rut80-1} were partly extended to the  
 subsystems consisting of $N'$ neighbouring atoms in the chain, $1<N'\ll N$. Starting from the same initial 
pure state $ |\psi(0)\rangle $, 
 the
time evolution of the reduced density operator $\rho^{(s)}(t)$  of the subsystem containing $N'$ atoms was 
calculated.
Qualitative behavior of $\rho^{(s)}(t)$  at large $t$ was shown to be  similar to that of the single-atom 
reduced density operator $\rho(k,t)$ described in \cite{rut80-1}. It should be noted, however, 
that calculation of the time-average in \cite{rut80-2} was not completely rigorous, 
but applied a certain reasonable conjecture, see equation (10) in \cite{rut80-2}.

The following  note is in order here. The quantum  chain of harmonic oscillators determined 
by Hamiltonian (\ref{HC}) has the 
complete set of $N$  commuting 
integrals 
of motion $\{n_l\}_{l=1}^N$ which characterize the number of phonons with given quasimomenta $\tau_l={2\pi l}/{N}$, 
$l=1,\dots,N$.  Integrability of model  (\ref{HC}) is the direct consequence of existence of these integrals 
of motion. 
The another side of the coin is that conservation of $N$ integrals of motion makes impossible the
complete thermalization of a typical initial state in this model.
However: 
\begin{itemize}
\item[(a)] An observer, which can see {\it {just one atom in the chain}}, can not distinguish phonon modes with
different quasimomenta $\tau_l$. By this reason, conservation of 
the phonon numbers $\{n_l\}_{l=1}^N$ in the whole quantum chain
does not conflict with the chain {\it {partial thermalization}}, which is 
registered by the observer watching at a
single atom.   
\item[(b)] Anyway, complete thermalization in model (\ref{HC}) could be still
 possible for some particular initial 
states of the chain.
\end{itemize}

The rest of this preprint is organized as follows. Section \ref{Sec1} describes in detail the background ideas
on thermalization in quantum systems, which stimulated me to do the model calculations described in \cite{rut80-1,rut80-2}. 
In a very brief form, these ideas were present already in the introductory parts of these two papers. 
Section \ref{Sec2} contains comments on the results and some calculation details of paper \cite{rut80-1}, 
which English translation is given in the 
Appendix. Section \ref{S3} contains new results illustrating the effect of the nonlinear 
interaction on the thermalization of the atom
energies  in the chain.
\section{Thermalization in isolated macroscopic quantum system \label{Sec1}}
The fundamental  question about the meaning and the mechanism of thermalization 
in a macroscopic isolated dynamical system can be divided into two parts:
\begin{itemize}
\item[(i)]
What is the precise meaning (microscopic definition) of the term 
'the thermal equilibrium state'
for the isolated macroscopic system?
\item[(ii)]
In what sense the isolated macroscopic system approaches to the equilibrium state
due to its dynamical evolution?
\end{itemize}

Though for classical systems, at least, partial answers to these questions could be found 
in the ergodic theory, 
the situation with the quantum systems was much less certain.
In the statistical mechanics textbooks  (for example, see \cite{Huang}), the equilibrium state 
of an isolated quantum system with the Hamiltonian ${H}$ and the energy  lying in the interval 
between $E$ and $E+\Delta$, ($\Delta\ll E$) 
is usually defined through the equal {\it a priori} probability
postulate, leading to the
microcanonical density operator $\rho_{\Delta}(E,  {H})$, 
\begin{eqnarray} \label{mc}
\rho
_{\Delta}(E,  {H}) =\frac{ g_{\Delta}(E,  {H}) }{{\rm Tr}\; g_{\Delta}(E,  {H})} ,\\
g
_{\Delta}(E,  {H}) =\theta({H}-E-
\Delta)-\theta({H}-E),
\end{eqnarray}
where $ {\rm Tr} \, {A}$ denotes the trace of the operator ${A}$,  and  $\theta(x)$  is the 
Heaviside step function. 

However, it is not 
easy to answer to the question (ii) with this conventional definition of the thermal equilibrium state. 
Really,  the matrix elements of the density operator $\rho(t)$  of the 
isolated system calculated in the
basis of its stationary states only change their phases with time and do not approach to
the matrix elements of the microcanonical  distribution (\ref{mc}).
Furthermore, evolution of the isolated system from some  non-equilibrium pure state $ |\psi(0)\rangle $
is described by the unitary  operator ${U}(t)=\exp(-i t {H})$, 
$|\psi(t)\rangle={U}(t) \,|\psi(0)\rangle$. Corresponding density operator 
$ {\rho}(t) =U(t)\,{\rho}(0)\,U^{-1}(t)$ being the orthogonal  projector operator at $t=0$, 
\begin{equation}\label{nes0}
{\rho}(0)=|\psi(0)\rangle\langle\psi(0)|,
\end{equation}
remains, of course, to be the orthogonal projector operator at all $t\ne0$,
\begin{equation} \label{nes}
{\rho}(t) =|\psi(t)\rangle\langle\psi(t)|.
\end{equation} 
Since the microcanonical density operator (\ref{mc})
is not a projector operator,  the projector-type density operator ${\rho}(t)$ can not 
approach to it in some simple
sense, say, in the strong or weak operator topology. 

The common way used in the textbooks to avoid this problem, is to remind that
the notion of the isolated system is a physical idealization, and in reality,
all systems interact with their surrounding. However, though this statement is undoubtedly correct,
one could hardly believe, that it is the true origin of thermalization, 
and that the latter can not be understood within the concept of the isolated quantum system.

If only isolated systems are concerned, the possible way to answer question (ii) is 
to restrict the set 'physically observable' operators ${A}_i$, to some family $ {\mathcal A} $ 
and to claim,  that  only for these operators $ A_i\in{\mathcal A} $ one should expect approaching of the 
quantum expectation values $ \langle A(t)\rangle $  to the their (microcanonical) equilibrium values:
\[
\langle A(t)\rangle={\rm Tr} \,[ {A}_i\,\cdot {\rho}(t) ] \leadsto  {\rm Tr} 
\,[ {A}_i\,\cdot \rho_{\Delta}(E,  {H})], 
\]
with subsequent fluctuations.\begin{footnote}
{
Notation 
 $f(t) \leadsto C$
will be used to indicate that the value of the time-dependent quantity $f(t)$ 
is close (equals in the thermodynamic limit) to $C$ at almost any moment $t$.
} 
\end{footnote}
This idea was realized by von Neumann in his Quantum Ergodic Theorem \cite{Neum29,Neumann}. 
In this theorem, operators $A_i$ were treated by von Neumann as operators of 'the 
macroscopic observables', 
and were assumed to commute with each other, \newline
$
[A_i, A_j]=0, \quad {\rm for} \; A_i,A_j\in {\mathcal  A}.
$

A deep analysis of  foundations of the statistical mechanics
 was given by Landau and Lifshitz in their Course 
of Theoretical Physics. They define the thermal equilibrium in the following
 way (see Pages 6 and  2 in  \cite{LL80}):
{\Definition \label{LL} . 
If a closed  macroscopic system is in a state such that in any {\uwave {macroscopic} }
subsystem the {\uwave{macroscopic} } physical quantities are to a high degree of 
accuracy equal to their mean values, the system is said to be in a state of 
statistical equilibrium (or thermodynamic or thermal equilibrium).}

\noindent
The subsystems here  are implied to be small 
compared with the whole system and 
weakly interacting with surrounding. Such subsystems will be called as 'allowed subsystems'
in the sequel.

In the above definition, thermalization behavior is required for the reduced family 
of 'the macroscopic observables', which relate to the macroscopic subsystems of the isolated system.

Taking definition \ref{LL} as a starting point,  one could arrive to a nice physical scenario,
in which thermalization in the isolated macroscopic quantum system  could result solely 
from its quantum mechanical evolution.
The first step was to drop two underlined words '{\uwave {\it {macroscopic}}}' in the above definition.
This extended the family ${\mathcal A}$ of 'thermalizable' physical observable to
 {\it all} physical observables, which characterize the state
of {\it allowed subsystems}, either macroscopic or not.  
Then, definition \ref{LL} becomes equivalent to the following one:

{\Definition \label{q} 
\cite{rut80-2}.
%
 A  closed   macroscopic  system  with  Hamiltonian  H  at the time moment t 
is  in the  state  of  equilibrium  if any  sufficiently  small,  possibly  
macroscopic  subsystem  of it that  is weakly 
coupled  to the  surroundings  has  the  Gibbs  distribution.   }

\noindent
Let us 
denote by ${\alpha}$ some subdivision of our isolated macroscopic quantum system 
into the allowed subsystem $\{s,\alpha\}$ with the Hamiltonian $H_{s,\alpha}$,
 and the rest of the system  $\{r,\alpha\}$,
 which plays the
role of the thermal bath. 
The Hilbert space $ {\mathcal L} $ of the whole system can be decomposed 
into the tensor product
$
{\mathcal L}= {\mathcal L}_{s,\alpha} \otimes {\mathcal L}_{r,\alpha}
$
of the Hilbert spaces ${\mathcal L}_{s,\alpha}$ and ${\mathcal L}_{r,\alpha}$ 
corresponding to the subsystem,  and its surrounding,
respectively. 

{\Definition \label{part} {An isolated macroscopic quantum system will be called to 
be in the (partial) thermal equilibrium with respect to the allowed subsystem $\{s,\alpha\}$,
if the reduced density operators $\rho^{(s,\alpha)}(t)$ of   this subsystem 
is to a high accuracy equal to the canonical density operators  $\rho_G^{(s,\alpha)}$ 
given by the Gibbs formula 
\begin{eqnarray} \label{G}
 \rho_G^{(s,\alpha)} = Z_{s,\alpha}^{-1} \,{ \exp [-\beta(\alpha) \, H_{s,\alpha}] } ,\\
Z_{s,\alpha} ={ {\rm Tr }^{(s,\alpha)} \,\exp [-\beta(\alpha) \, H_{s,\alpha}]}, \nonumber
\end{eqnarray}
where
$ {\rm Tr }^{(s,\alpha)} $ denotes the trace over the Hilbert space corresponding to the 
quantum states of the 
subsystem $\{s,\alpha\}$, and $\beta(\alpha)$ is the inverse temperature which could 
depend on the subsystem $\{s,\alpha\}$.}}

\noindent
Note, that the reduced density operator of the subsystem  $ \rho^{(s,\alpha)}(t) $ 
is related with the  density operator $ \rho(t) $ of the whole system as
\begin{equation}\label{rd}
\rho^{(s,\alpha)}(t)= {\rm Tr }^{(r,\alpha)} \, \rho(t),
\end{equation}
where ${\rm Tr }^{(r,\alpha)}$ denotes the trace over the thermal bath space ${\mathcal L}_{r,\alpha}$.

It is clear, that the macroscopic quantum system stays in the  global thermal equilibrium in the sense of
 definition \ref{q}, 
if and only if it is in the partial thermal equilibrium with respect to all its allowed subsystems $\{s,\alpha\}$
with the same inverse temperature $\beta$.

Definition \ref{q} has several important advantages.
First, it does not refer to the notion of the microcanonical ensemble and to the equal {\it a priori} 
probability postulate. 
Second, with { {definition~\ref{q}}}, the answer to the question (ii) becomes straightforward and leads to
the natural treatment of the  thermalization phenomenon in quantum systems.
{\proposition \label{t} {
Thermalization in the macroscopic isolated quantum system  means, that after the relaxation time, 
the matrix elements 
of the reduced
density operators  $ \rho^{(s,\alpha)}(t)$ of all its allowed subsystems $\{s,\alpha\}$ in the basis of 
their
stationary states $| \Phi_n^{(s,\alpha) } \rangle $ should approach (up to further fluctuations) to those
of the canonical density operators (\ref{G}),
\begin{eqnarray} \label{G1}
&&\langle \Phi_n^{(s,\alpha)} |\rho^{(s,\alpha)}(t)| \Phi_{n'}^{(s,\alpha) } \rangle \leadsto 
 \,Z_{s,\alpha}^{-1}\, 
\exp(-\beta {\mathcal E}_n^{(s,\alpha)})\,\delta_{n,n'} ,\\
&&  H^{(s,\alpha)} | \Phi_n^{(s,\alpha) } \rangle ={\mathcal E}_n^{(s,\alpha)} \,| \Phi_n^{(s,\alpha) } 
\rangle, \nonumber\\
&&{\rm where  } \quad  Z_{s,\alpha}=\sum_n  \exp(-\beta {\mathcal E}_n^{(s,\alpha)}), \nonumber
\end{eqnarray}
and ${\mathcal E}_n^{(s,\alpha)}$, $n=1,2,\dots$ is the energy spectrum of the subsystem $\{s,\alpha\}$.
 }}

\noindent
Third, one can easily see, that the problem mentioned in the beginning of this Section
does not arise with definition \ref{t}.
Really, if the whole system is in a pure state described by some  projector  density operator (\ref{nes}), 
the subsystem reduced density operator (\ref{rd})   can correspond to a mixed state due to 
the quantum entanglement, and this mixed state can approach to the equilibrium  Gibbs state 
(\ref{G}). 

If a macroscopic isolated quantum system possesses the property of thermalization, one should expect, that 
a typical 
initial state $|\psi(0)\rangle$ having macroscopically  well determined total energy would approach 
in its unitary evolution to the equilibrium state,
understood  according to definition \ref{q} and then remain in this state almost at  any time.

It is tempting to take the initial state $|\psi(0)\rangle $ of the whole system as a single
stationary state $ |n\rangle $ of its Hamiltonian, $H |n\rangle =E_n |n\rangle $.
 Then, the previous statement naturally leads to the 
ETH \cite{Deutsch91,Sred94}, 
 which can be formulated as follows:   
{\proposition \label{ETH}{
In a thermalizable isolated quantum system, the huge majority of individual stationary states 
$ |n\rangle $ of the whole system represent 
the thermal equilibrium state. This means, that
\begin{equation} 
{\rm Tr }^{(r,\alpha)}|n\rangle \langle n| \,=Z_{s,\alpha}^{-1} \,{ \exp (-\beta \, H_{s,\alpha}) },
\end{equation}
for all allowed subsystems $\{s,\alpha\}$, and almost all stationary states $ |n\rangle $.}}

Though the described above heuristic scenario of thermalization looked rather reasonable and attractive for me
in 1979, I had no idea how to put it on a more firm basis for a generic isolated quantum system. So, 
it was natural to try to test this scenario on some simple model of interacting particles.    
Such a test was partly realized in \cite{rut80-1,rut80-2} for the model of the quantum chain of harmonic oscillators. 
  
It should be noted, that definition \ref{q} provided  the basis of the described above treatment of thermalization.   
It implies that the family ${\mathcal A}$ of thermalizable 
observables can be associated with all observables which characterize the states of allowed subsystems. 
Approach to the thermalization problem emphasizing the role of the canonical ensemble for the subsystem state as 
the main indication 
of thermal equilibrium in quantum systems,\begin{footnote}{ The  origin of this approach 
comes back to Erwin Schr{\"o}dinger \cite{Schr52}, see discussion in \cite{Leb06}. }\end{footnote}
was developed later by several groups. 
Important results along this direction were 
obtained by Tasaki \cite{Tas98}, Popescu, Short, and Winter \cite{Po06},
Linden {\it et al.} \cite{Lin09}, 
and by Goldstein {\it et al.}  \cite{Leb06,Leb10a}.
\section{Comments on the results of paper \cite{rut80-1} \label{Sec2}}
 In the normal mode representation, the quantum chain of harmonic oscillators reduces to the free phonon gas. 
  As it was mentioned in the Introduction, conservation of phonon numbers 
$\{n_\tau\}_{\tau=1}^{2\pi/N}$ makes impossible complete thermalization of a typical initial state 
$|\psi(0)\rangle$ in this model. However, the quantum measurements performed on  one atom in the chain
does not allow one to  distinguish different phonon modes. So, it is
natural to  expect, that for model (\ref{HC}),
the
weak versions 
of the thermalization property (proposition \ref{t}) and of the ETH (proposition \ref{ETH}) are 
still valid, in which Equations (\ref{G}) and (\ref{G1}) are required only for the subsystems 
 $\{s,k\}$ consisting of one $k$-th atom in the chain.\begin{footnote} 
{Note, that due to the requirement $\Omega\ll\Omega_0$, each atom in the chain weakly interacts 
with its surrounding and, therefore,  represents an allowed subsystem by itself. }
\end{footnote}
Paper \cite{rut80-1}, which English translation is given in Appendix,  
was dedicated to verification of this statement. 
\subsection{Eigenstate thermalization for free bosons}
The Eigenstate Thermalization Hypothesis (see proposition \ref{ETH})  was the important part of the 
described above heuristic scenario of thermalization, which provided motivation for \cite{rut80-1,rut80-2}. 
Note, that the ETH was not formulated explicitly in \cite{rut80-1,rut80-2}, this has been done later by 
Deutsch \cite{Deutsch91} and Srednicki \cite{Sred94}. However, a weak form of the  ETH for model (\ref{HC}) 
 was proved in \cite{rut80-1}. Namely, 
equation \eqref{Tr1} validates the ETH (proposition \ref{ETH}) in model (\ref{HC}) for 
the {\it single-atom  subsystems}. It was conjectured also in \cite{rut80-1}, that,
"perhaps, this result holds to some extent in other dynamical systems".

Parameter $\bar{n}=   {\mathfrak N}  /N$ in \eqref{Tr1}
is related with the inverse temperature
$\beta$ by the Plank's formula
\begin{equation}
\bar{n}=\frac{1}{\exp(\beta\, \Omega_0)-1},
\end{equation}
and $ {\mathfrak N} =\sum_{l=1}^N n_l$.
In the case $\bar{n}<1$, equation \eqref{Tr1} can be easily obtained from \eqref{Tr}
which, in turn,  follows from the results of Appendix 1  of paper \cite{rut80-2}.

To prove \eqref{Tr1}, let us rewrite  \eqref{Tr} for $n=n'$ in the explicit form
\begin{equation} \label{tr}
{\rm Tr}^{(r)}|n\rangle\langle n|= \sum_{J=0}^{\mathfrak N}|J\rangle\langle J|
\sum_{P=J}^{\mathfrak N} \frac{P!^2(-1)^{P-J}}{N^P J! (P-J)!}
\sum_{\sum_l m_l=P}\,
\prod_{l=1}^N
\frac{{n_l!}}{(n_l-m_l)!\,m_l!^2} ,
\end{equation}
where $m$ denotes the set of integer numbers $\{m_l\}_{l=1}^N$,
which lie in the interval $0\le m_l\le n_l$. Denote by 
$N_1$ the number of nonzero integers in the set $\{n_l\}_{l=1}^N$.  
For fixed $P>0$ and macroscopic $N_1$, the leading contribution to the sum 
\begin{equation} \label{su}
S_2(n_1,\dots,n_N;P) =  \sum_{\sum_l m_l=P}\,
\prod_{l=1}^N
\frac{{n_l!}}{(n_l-m_l)!\,m_l!^2}
\end{equation}
comes from configurations $\{m_l\}_{l=1}^N$, in which  $m_l$ takes values either $0$, or $1$. 
Contribution of all the rest terms is smaller by the factor $N_1^{-1}$.
The same is valid also for the sum
 \begin{eqnarray} \label{su1}
S_1(n_1,\dots,n_N;P) =\sum_{\sum_l m_l=P}\,
\prod_{l=1}^N
\frac{{n_l!}}{(n_l-m_l)!\,m_l!}
= \frac{1}{P!}{(\partial_z)^P}|_{z=0} \prod_{l=1}^N (1+z)^{n_l} = \\
\nonumber
\frac{1}{P!}{(\partial_z)^P}|_{z=0} (1+z)^{\mathfrak N}    = \frac{ { \mathfrak N} ! }{P!({ \mathfrak N}-P)!}.
\end{eqnarray}
Therefore, 
\begin{eqnarray} \label{su2}
S_2(n_1,\dots,n_N;P) =S_1(n_1,\dots,n_N;P) [1+O(N_1^{-1})]  =\\ \nonumber
\frac{ { \mathfrak N} ! }{P!({ \mathfrak N}-P)!}
[1+O(N_1^{-1})] =
\frac{ { \mathfrak N}^P }{P!}
[1+O(N_1^{-1})],
\end{eqnarray}
at fixed $P$ and ${\mathfrak N}\sim N_1\sim N\to\infty$. Formal substitution of \eqref{su2} into
\eqref{tr} yields
\begin{eqnarray} \label{tr1}
{\rm Tr}^{(r)}|n\rangle\langle n|= \sum_{J=0}^\infty  |J\rangle\langle J|
\sum_{P=J}^{\infty} \frac{P!(-1)^{P-J}\, {\bar n}^P }{ J! (P-J)!} [1+O(N_1^{-1})] =\\\nonumber
\,[1+O(N_1^{-1})]\, \sum_{J=0}^\infty  \frac{{\bar n}^J}{(1+{\bar n})^{J+1}}\, |J\rangle\langle J|,
\end{eqnarray}
in agreement with \eqref{Tr1}.
If ${\bar n}<1$, the above formal manipulations can be 
justified, since the infinite series in $P$ in \eqref{tr1} converges uniformly in $N\to\infty$.

At ${\bar n}\ge 1$, the series in $P$ in \eqref{tr1} diverges. Nevertheless, it is also possible 
to prove \eqref{Tr1} at ${\bar n}\ge 1$ directly from \eqref{Tr}, though the proof is more complicated, than 
at  ${\bar n}< 1$. On the other hand,  a much easer indirect way to confirm \eqref{Tr1} exists,
which is suitable for all positive $\bar{n}$. Namely,
one can check that the 
quantum expectation values of the operator $b_0^{\dagger J_1}b_0^{J_2}$ 
calculated with the both sides of this equation lead to the same results
for any integer $J_1,J_2\ge0$.
{\proposition
{ 
Let $\{ a_l\}_{l=1}^N$ and $\{b_k\}_{k=1}^N$ are the two sets of $N$ bosonic annihilation operators,
which are related by the linear transformation 
\begin{equation}\label{b}
b_k=\sum_{l=1}^N A_{kl}(N) \,a_l, 
\end{equation}
with the unitary
$N$-dependent matrix $A_{kl}(N) $, and 
\begin{equation}\label{n1}
|n_1,\ldots, n_N\rangle=(n_1!\ldots n_N!)^{-1/2}(a_1^\dagger)^{n_1}\ldots
(a_N^\dagger)^{n_N} |0\rangle,
\end{equation}
denotes the basis associated with the first set of the bosonic operators. For each $N$, let us choose a basis
vector $|n_1(N),\ldots,n_N(N)\rangle$ in such a way, that
\begin{equation} \label{con}
 |A_{kl}(N)|^2 n_l(N)<{C}/{N} \quad  \textrm{
for all } k,l\in[1,N], \textrm{  with some fixed  } C>0,
\end{equation}
and
the limits 
\begin{equation} \label{con1}
{\bar{n}}(k)= \lim_{N\to\infty} \sum_{l=1}^N |A_{kl}(N)|^2 n_l(N),   
\end{equation}
exist for given $k$.
Then
\begin{equation} \label{cor}
\lim_{N\to\infty}  \langle n_N(N),\ldots,n_1(N)|  b_k^{\dagger M_1}b_k^{M_2}  |n_1(N),\ldots,n_N(N)
\rangle  =   M_1 !\,[ {\bar{n}}(k) ]^{M_1}\, \delta_{M_1 M_2}.
\end{equation}
}}
\begin{proof}
First, it is clear, that operator $b_k^{\dagger M_1}b_k^{M_2}$ with $M_1\ne M_2$ has zero 
diagonal matrix elements in the basis \eqref{n1}, since this operator changes the number of bosons.
So, it is sufficient to put $M_1=M_2\equiv M$ in \eqref{cor}. Then, 
substitution of \eqref{b} into the matrix element in the left-hand side of this equation followed by
straightforward calculations yields
\begin{eqnarray}  \nonumber
&&\langle n_N,\ldots,n_1|  b_k^{\dagger M}b_k^{M}  |n_1,\ldots,n_N\rangle  =\\   \nonumber
&&\sum_{ l_1+\ldots+l_N=M}\frac{M!^2}{l_1!^2\ldots l_N!^2}|A_{k l_1}\ldots A_{k l_N}|^2 
\langle n_N,\ldots,n_1| a_N^{\dagger l_N} \ldots a_1^{\dagger l_1}a_1^{l_1}\ldots a_N^{l_N} 
|n_1,\ldots,n_N\rangle=\\   \nonumber
&&\sum_{ l_1+\ldots+l_N=M}\frac{M!^2}{l_1!^2\ldots l_N!^2}|A_{k l_1}\ldots A_{k l_N}|^2 
\frac{n_1!\ldots  n_N!}{(n_1-l_1)!\ldots (n_N-l_N)!}=\\
&&M!\,\frac{\partial^M}{\partial t^M} \bigg|_{t=0} \prod_{l=1}^N L_{n_l}(-t\,|A_{k l}|^2 ),  \label{la}
\end{eqnarray}
where $L_n(z)$ is the Laguerre polynomial,
\[
L_n(z)=\sum_{l=0}^n \frac{n!(-z)^l}{l!^2(n-l)!}.
\]
The explicit form of \eqref{la}  read as 
\begin{eqnarray*}
&&\langle n_N,\ldots,n_1|b_k^{\dagger}b_k|n_1,\ldots,n_N\rangle  =  \sum_{l=1}^N |A_{kl}|^2 n_l,\\
&&\langle n_N,\ldots,n_1|b_k^{\dagger 2}b_k^2|n_1,\ldots,n_N\rangle  = 
2\left[ \sum_{l=1}^N |A_{kl}|^2 n_l\right]^2-
\sum_{l=1}^N |A_{kl}|^4 ( n_l +n_l^2) ,\nonumber\\
&&\langle n_N,\ldots,n_1|b_k^{\dagger 3}b_k^3|n_1,\ldots,n_N\rangle  =
3!\left[ \sum_{l=1}^N |A_{kl}|^2 n_l\right]^3+
\sum_{l=1}^N |A_{kl}|^6 (2 n_l+6 n_l^2+4 n_l^3)- \\
&&9 \sum_{l=1}^N |A_{kl}|^4 \sum_{m=1}^N ( n_l\,n_m +  n_l^2),\\
&&\langle n_N,\ldots,n_1|b_k^{\dagger M}b_k^M|n_1,\ldots,n_N\rangle =
  M!\left[ \sum_{l=1}^N |A_{kl}|^2 n_l\right]^M+\ldots
\end{eqnarray*} 
It is easy to understand, that only the first term in the right-hand side of the above relations survives in the 
thermodynamic limit $N\to\infty$ due to \eqref{con} and \eqref{con1}.
\end{proof}
Let us associate with each pair of operators $b_k^\dagger, b_k$ the 'single-atom subsystems' $\{s,k\}$
with the Hamiltonians
$h_k^{(s)}=\epsilon_k\, b_k^{\dagger}b_k$. Suppose now, that each such atom is in the 
thermal equilibrium state characterized 
by the $k$-dependent inverse temperature 
\begin{equation}\label{bet}
\beta_k=\epsilon_k^{-1}\log{\frac{1+{\bar{n}}(k)}{{\bar{n}}(k)}}. 
\end{equation}
This means, that the reduced density operator of the $k$-th atom has the Gibbs form,
\begin{equation}\label{Gi}
\rho_G(k)=\frac{\exp{[-\beta(k) h_k^{(s)}]}}{Z_k^{(s)}}
=\sum_{J=0}^\infty \frac{[ {\bar{n}}(k) ]^J}{[1+ {\bar{n}}(k) ]^{J+1}}\,\frac{1}{J!}\,\,
b_k^{\dagger J}|0\rangle\langle0|b_k^J.
\end{equation}
 Then,  for arbitrary natural 
$M_1,M_2$, one can easily obtain the
expectation value of the product $b_k^{\dagger M_1}b_k^{M_2}$  in this state
\begin{equation}
{\rm Tr}^{(s,k)}\left[  b_k^{\dagger M_1}b_k^{M_2} \, \rho_G(k)  \right] = 
M_1 !\,[ {\bar{n}}(k) ]^{M_1}\, \delta_{M_1 M_2},
\end{equation} 
which coincides with the right-hand side of \eqref{cor}. 

Thus, at large $N$, the huge majority 
of basis states \eqref{n1} correspond to the partial equilibrium states with respect 
to the single-atom subsystems $\{s,k\}$ in the sense of definition \ref{part}. The
inverse temperature $\beta_k$ of the $k$-th atom, generally speaking, depends on $k$, 
and is given by equations \eqref{bet}, \eqref{con1}.

Let us return to the chain of harmonic oscillator \eqref{HC}. For the 
 bosonic annihilation operators $b_k$ and $a_l$ determined according to
\eqref{ab}, the matrix $A_{kl}(N)$  introduced in \eqref{b} reduces at $\Omega=0$ to 
the simple form $A_{kl}= N^{-1/2}\exp{(2\pi i \,l/N)}$, and $\epsilon_k=\Omega_0$.
 Parameter $\bar{n}(k)$ defined by 
\eqref{con1} reduces to the average number of phonons in the chain $\bar{n}=N^{-1}\sum_{l=1}^N n_l$ 
and becomes $k$-independent, together with the inverse temperature \eqref{bet}. 
This proves \eqref{Tr1} for all $\bar{n}>0$.
\subsection{Time evolution of the single-atom density operators}
Relaxation of the macroscopic chain of harmonic oscillators from a nonequilibrium nonuniform initial state
$|\psi(0)\rangle =({\mathfrak N}!)^{-1/2}\, b_0^{\dagger {\mathfrak N}}|0\rangle$ was studied in Section 2
of \cite{rut80-1}. First, the time evolution of the chain quantum state $|\psi(t)\rangle$ was determined in 
the leading order in the interaction constant $\Omega\to0$. Then, representation (\ref{rho}) was obtained for the 
density operators $\rho(k,t)$ of the $k$-th atom at time $t$, which is asymptotically exact at $\Omega\to 0$.
The full form of this expression reads as
\begin{equation}\label{rhoa}
\rho(k,t) =\sum_{J=0}^{\mathfrak N}|J\rangle\langle J| \frac{{\mathfrak N}!}{J!({\mathfrak N}-J)!}
\left[\frac{r(k,t)}{N}\right]^{2J}\left\{1-\left[\frac{r(k,t)}{N}\right]^{2}\right\}^{{\mathfrak N}-J},
\end{equation}
with
\begin{equation} \label{r2}
r(k,t) = \bigg| \sum_{l=1}^N \exp{[ i(\tau_l k - \omega_{\tau_l} t)]} \bigg| .
\end{equation}
Remind, that $\tau_l=2\pi l/N$ and $\omega_{\tau}$ are the phonon quasimomentum and the energy, respectively, 
\begin{eqnarray}\label{ome}
\omega_\tau =[\Omega_0^2+4 \Omega^2 \sin^2(\tau/2)]^{1/2}=\Omega_0+\frac{ \Omega^2}{\Omega_0}(1-\cos \tau)+O(\Omega^4).  
 \end{eqnarray} 

\begin{figure}[htb]
\centering
\includegraphics[width=\linewidth, angle=00]{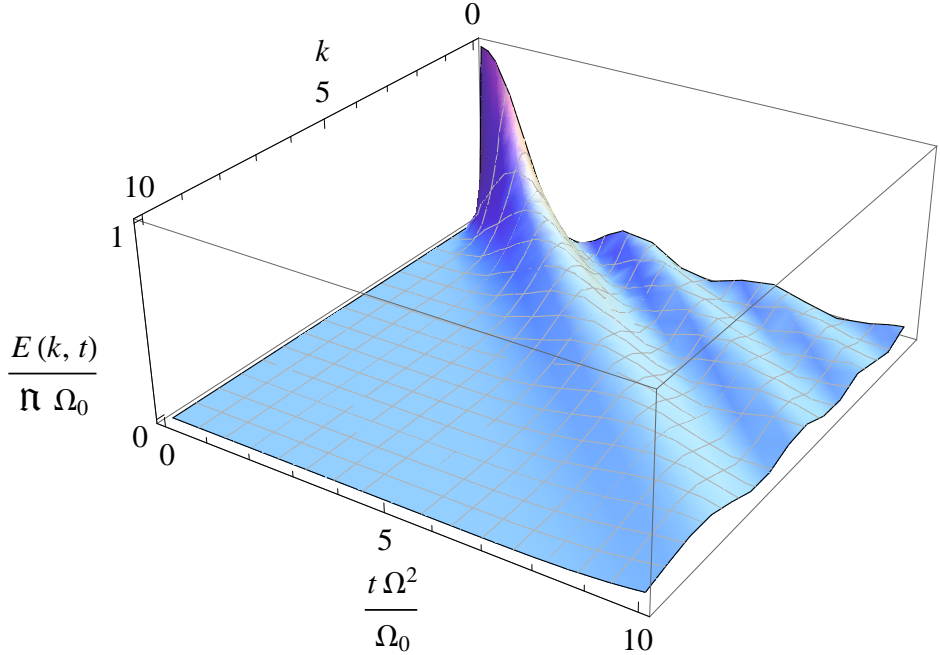}
\caption{\label{rel} Space and time dependence of the energy  localized on the 
$k$-th atom  normalized to 
the total energy in the chain,  
 $E(k,t)/({\mathfrak N}\, \Omega_0)=[J_k(t\Omega^2/\Omega_0)]^2$.} 
\end{figure}

Parameter $r^2(k,t)$ is proportional to the energy $ E(k,t) $, which is   located at the $k$-th atom at the time moment $t$:
\begin{equation}\label{en}
E(k,t)= \Omega_0 \,\langle\psi(t)| b_k^\dagger b_k |\psi(t)\rangle= 
\Omega_0 \,{\rm Tr}^{(s,k)}[\rho(k,t)\,b_k^\dagger b_k]=
{ \mathfrak N}\, \Omega_0 \,r^2(k,t)  /{N^2} .
\end{equation}
At small enough $t$, the sum in (\ref{r2}) can be replaced by the integral. This allows one to 
express the energy $E(k,t)$ normalized to the total energy in the chain in terms of the cylindric  Bessel function $J_k(z)$,
\begin{eqnarray}\nonumber
\frac{E(k,t)}{{\mathfrak N}\, \Omega_0}= \frac{r^2(k,t)}{N^2}  \approx   \bigg|\int_0^{2\pi}\frac{d\tau}{2\pi}
e^{ i(\tau k - \omega_{\tau} t)} \bigg|^2 \approx  \bigg|\int_0^{2\pi}\frac{d\tau}{2\pi}
\exp\left[{ i\left(\tau k + \frac{\Omega^2 t  }{\Omega_0} \cos \tau \right)}\right] \bigg|^2= \\
\left[ J_k(t\,\Omega^2/\Omega_0)\right]^2. \label{Bes}
\end{eqnarray}
The plot of this normalized energy versus $k$ and $t$ is shown in Figure \ref{rel}.
It is clear from (\ref{En}) and Figure \ref{rel}, that the energy perturbation spreads along the chain with the 
velocity $v= \Omega^2/\Omega_0 $, which is just  the maximum group velocity of phonons 
\begin{eqnarray*}
&& v_{ph}(\tau) =\frac{d \omega_\tau}{d\tau}
= (\Omega^2/\Omega_0)\, \sin \tau + O(\Omega^4), \\
&&v= \max_\tau v_{ph}(\tau)=v_{ph}(\pi/2). 
\end{eqnarray*}
This feature of the relaxation 
dynamics known as 'the horizon effect'  is typical in the quantum quench phenomenon \cite{CC07}.
\subsection{Dispersion of the density operators \label{ddo}}
\begin{figure}[htb]
\centering
\includegraphics[width=.8\linewidth, angle=00]{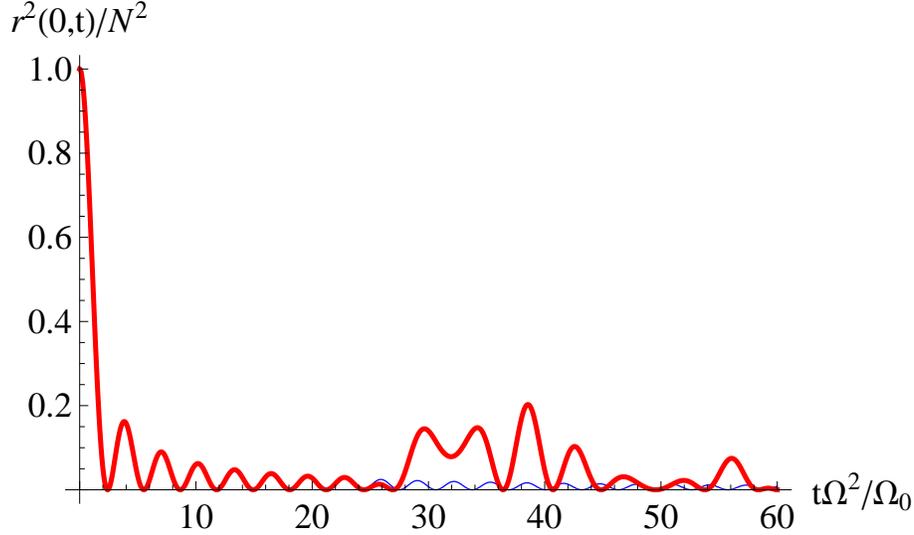}
\caption{\label{R2} The red line shows the time dependence of the quantity $r^2(0,t)/N^2$ 
defined by (\ref{r2}), which 
is proportional to the energy $E(0,t)$ localized on the 
$0$-th atom, $E(0,t)/( { \mathfrak N}\, \Omega_0)=\,r^2(0,t)  /{N^2}$. The blue line plots  
$[J_0(t\,\Omega^2/\Omega_0)]^2$. The model parameter values are taken as
$N=30,\; \Omega^2/\Omega_0=0.01$. Change of the evolution regime takes place at $t\simeq t_r, \;
t_r\,\Omega^2/\Omega_0  =30 $.
} 
\end{figure}

Now let us turn to the dynamics on the long time scale.  When time $t$ becomes  larger 
then the relaxation time $t_r=N/v= N \Omega_0/\Omega^2$, the energy perturbation reaches
all atoms in the chain, all exponential terms in the sum in right-hand side of (\ref{r2}) completely dephase
from each other, and  
this sum cannot be approximated by the integral, as it was done in  (\ref{Bes}). 
This leads to the qualitative change of  of the chain evolution character from regular at  $ t  \lesssim t_r$, 
to the 'stochastic' regime at $t\gtrsim t_r$, which is illustrated in Figure \ref{R2}.

For the atom reduced density operators averaged  over the infinite time interval, the 
canonical Gibbs distribution (\ref{r6}) with the site-number independent temperature  
was obtained in \cite{rut80-1}. However, 
derivation of this result was
incorrect, since it was based there on the erroneous equation (A.14a) in Page \pageref{r5}, 
see the footnote in this page.  The fallacy in this equation came from the implicit and 
mistaken assumption, that the energy levels $E_n$ of the whole system are non-degenerate,
\begin{equation}\label{nd}
E_n\ne E_{n'}, \quad{\rm if}\; n\ne n'.
\end{equation}
Really, the energy levels (measured from the ground state energy) in model (\ref{HC}) are given by  
\begin{equation} \label{ele}
E_n=(n\omega)\equiv \sum_{\tau\in {\mathcal B}} n_\tau\,\omega_\tau,
\end{equation}
where summation in the quasimomentum 
$\tau$ runs  over the Brillouin zone, 
\newline ${\mathcal B}=\frac{2\pi}{N}\cdot ({\mathbb Z}\mod{N}$). Due to the
 the mirror symmetry $\omega_\tau=\omega_{-\tau}$, 
of the phonon dispersion law \eqref{ome}, this 
energy spectrum can be rewritten in the form 
\begin{equation} \label{ele1}
E_n=n_0\,\omega_0+ \frac{n_\pi\,\omega_\pi}{2}\left[1+(-1)^N\right]+
\sum_{\substack{\tau\in {\mathcal B}\\ 0<\tau<\pi}} (n_\tau+n_{-\tau})\,\omega_\tau,
\end{equation}
which indicates the strong degeneracy of the energy levels $E_n$ in model \eqref{HC}, 
in contrast to \eqref{nd}.
Accordingly, one gets for the time average $\overline{\exp[it(E_{n'}-E_n)]}$:
\begin{equation} \label{avv}
\overline{\exp[it(\omega(n'-n))]}
=\begin{cases}
\delta_{n_0, n_0'} \,\delta_{n_\pi, n_\pi'}\prod_{\substack{\tau\in {\mathcal B}\\ 0<\tau<\pi}} 
\delta_{ n_\tau+n_{-\tau} ,n_\tau'+n_{-\tau}'},&{ \rm for\; even}\; N, \\
\delta_{n_0, n_0'}\prod_{\substack{\tau\in {\mathcal B}\\ 0<\tau<\pi}} 
\delta_{ n_\tau+n_{-\tau} ,n_\tau'+n_{-\tau}'},&\,{ \rm for\; odd}\; N,
\end{cases}
\end{equation}
instead of equation (A.14a). It turns out, however, that this correction does not 
change the final result (\ref{r6}) for the time averages of the reduced density operators
almost for all atoms in the chain.  
{\proposition \label{av} 
In the thermodynamic limit, the time averages of the atom reduced density operators 
$\rho(k,t)$ determined by equation \eqref{rho} [or, eqivalently, by
equations (\ref{rhoa}), \eqref{r2}] are given by the  canonical 
Gibbs distributions with two different temperatures,
\begin{equation} \label{avr}
\lim_{\substack {N\to\infty\\
 {\mathfrak N}/N=\bar{n}}}  \overline{\rho(k,t)} =\begin{cases}
 \sum_{J=0}^\infty  \frac{(2{\bar n})^J}{(1+2{\bar n})^{J+1}}\, |J\rangle\langle J|, & {\rm for }\;k=0,\\
 \sum_{J=0}^\infty  \frac{(2{\bar n})^J}{(1+2{\bar n})^{J+1}}\, |J\rangle\langle J|, &{\rm for\;} k=N/2\; {\rm and \; even\;} N,\\
 \sum_{J=0}^\infty  \frac{{\bar n}^J}{(1+{\bar n})^{J+1}}\, |J\rangle\langle J|, &{\rm otherwise},\\
\end{cases}
\end{equation}
where the time average is defined as
\begin{equation}\label{timeav}
\overline{ \rho(k,t) }\equiv \lim_{T\to\infty}\frac{1}{T}\,\int_{0}^T dt\,\rho(k,t).
\end{equation}}
Equation (\ref{avr}) corrects equation (\ref{r6}) for   $k= 0$, and $k=N/2$.
\begin{proof}
 will be given only for the case of even $N$, since extension to the case 
of odd $N$ is straightforward.

Let us introduce the $(\frac{N}{2}+1)$-dimensional phase space with the points 
$ \phi=\{\phi_\tau\} $, where $0\le \phi_\tau<2\pi$, and $\tau=\tau(l)=2\pi l/N$, with
$l=0,1,\ldots, N/2$. Let us define the operator-valued
function $\rho(k,\phi)$ on this space which is obtained from
(\ref{r1}) by the replacement
\begin{eqnarray*}
t(n\omega)\to (n\phi)=n_0 \phi_0+ n_\pi \phi_\pi+\sum_{0<\tau<\pi}(n_\tau+n_{-\tau})\phi_\tau,\\
t(n'\omega)\to  (n'\phi)=n_0' \phi_0+ n_\pi' \phi_\pi+\sum_{0<\tau<\pi}(n_\tau'+n_{-\tau}')\phi_\tau, 
\end{eqnarray*}
and averaging over the phases $\{\phi_\tau \}$ of the $2\pi$-periodical functions $f(\phi)$,
\[
\langle f(\phi)\rangle_{\phi}=\int_0^{2\pi}f(\phi)  \prod_{l=0}^{N/2}   \frac{\phi_{\tau(l)}}{2\pi}.
\]
Then, it follows  from (\ref{avv}), that 
\begin{equation}
\overline{\exp[it(\omega(n'-n))]}=\langle\exp[i(\phi(n'-n))]\rangle_\phi. 
\end{equation}

Applying this equality to (\ref{r1}), one can conclude, that the time average of the operator
$\rho(k,t)$ is equal to the phase average of the operator $\rho(k,\phi)$.
Therefore, the time average of the operator (\ref{rhoa}) can be written as
\begin{equation}\label{R5}
\overline{\rho(k,t)} =\sum_{J=0}^{\mathfrak N}|J\rangle\langle J|\,\, C_{\mathfrak N}^J\,\,
\Big\langle[r(k,\phi)/N]^{2 J}\left[1-\left[r(k,\phi)/N\right]^2\right]^{{\mathfrak N}-J}
\Big\rangle_\phi,
\end{equation}
where
\begin{eqnarray*}
r(k,\phi)= \left|\sum_{l=0}^{N/2}w_l(k)\,\exp[i \phi_{\tau(l)}]\right|,\\
w_l(k)=
\begin{cases}
1, & {\rm for} \; l=0,\\
2 \cos[k\, \tau(l)],& {\rm for} \; l=1,\ldots,\frac{N}{2}-1,\\
(-1)^k, & {\rm for} \ l=\frac{N}{2}.
\end{cases}
\end{eqnarray*}
Accordingly, the matrix elements of the operator (\ref{R5}) in the basis 
$ |J\rangle  =(J!)^{-1/2}{b_k^{\dagger J}}|0\rangle$ take the form
\begin{eqnarray}\label{JJ}
&&\langle J'|\overline{\rho(k,t)}|J\rangle = \delta_{JJ'}\, \overline{\rho_J(k,t)} ,\\
&&\overline{\rho_J(k,t)}= \frac{{\mathfrak N}!}{J!({\mathfrak N}-J)!}\,\,
\int_0^\infty dr\, (r/N)^{2 J}\left[1-\left(r/N\right)^2\right]^{{\mathfrak N}-J}
p_N(k,r) ,\label{43}\\
&&p_N(k,r) = \langle \delta[r-r(k,\phi)]\rangle_{\phi}. \label{pN}
\end{eqnarray}
After rescaling of the integration variable $r=\sqrt{N}\, x$ 
in the right-hand side of (\ref{43}), the large-$N$ asymptotics of this equation can be written as
\begin{equation}\label{rh1}
\overline{\rho_J(k,t)}=
\int_0^\infty dx\, \frac{\bar{n}^J}{J!}\,\exp(-\bar{n}\,x^2)\, \sqrt{N}\,p_N(k,r)|_{r=x\sqrt{N}}+O(N^{-1}).
\end{equation}

Function $p_N(k,r)$ defined by (\ref{pN}) is the probability density to find the total length $r$
of the sum of $N/2+1$ vectors in the plane which have arbitrary directions and different lengths 
$|w_l(k)|, \quad l=0,\ldots,N/2$.
Straightforward calculations lead  to the 
exact representation of this probability density  in terms of the cylindrical Bessel function $J_0(z)$
[cf. equation \eqref{A16} in Appendix],
\begin{equation} \label{pNk}
 p \,_N(k,r) = r \int_0^\infty dv \,v \,J_0(v r ) \prod_{l=0}^{N/2} J_0[w_l(k)^2\, v].
\end{equation} 
At large $N$, the main contribution to the integral in the right-hand side of \eqref{pNk} 
comes from small $v$, since the integrand vanishes at 
 $v\gg N^{-1/2}$. So, in the thermodynamic limit   we get from (\ref{pNk}):
\begin{eqnarray} \nonumber
&& \sqrt{N}\,p_N(k,r)|_{r=x\sqrt{N}}  =N\, x\,  \int_0^\infty dv \,v \,J_0\left(v x\sqrt{N} \right) \exp\left[-\frac{N\,v^2}{4} Y(k,N)
\right][1+O(N^{-1})] =\\
&& \frac{2 x}{Y(k,N)} \exp[-x^2/Y(k,N)]+O(N^{-1}),\label{pNk1}
\end{eqnarray} 
where 
\begin{equation}\label{Sk}
Y(k,N)=\frac{1}{N}\sum_{l=0}^{N/2} w_l(k)^2=
\begin{cases}
2-\frac{2}{N}, & {\rm if \:}k=0,N/2,\\
1-\frac{2}{N}, & {\rm if \:}k=1,\ldots,\frac{N}{2}-1.
\end{cases}
\end{equation}
Substitution of (\ref{pNk1}) and (\ref{Sk}) into \eqref{rh1} yields finally to the result
\begin{equation}
\overline{\rho_J(k,t)}=\frac{ [\bar{n}\,Y(k,N)]^J }{[1+\bar{n}\,Y(k,N)]^{J+1}}+ O(N^{-1}) =
\begin{cases}
\frac{[ 2\bar{n} ]^J}{(1+2\bar{n})^{J+1}} + O(N^{-1}),& {\rm if \:}k=0,N/2,\\
\frac{ \bar{n}^J}{(1+\bar{n})^{J+1}} + O(N^{-1}),& {\rm if \:}k=1,\ldots,\frac{N}{2}-1,
\end{cases}
\end{equation}
which, together with (\ref{JJ}) proves \eqref{avr}.
\end{proof}

Let us return now to the time evolution of the atom reduced density operators (\ref{rho}) at large 
$t\gtrsim t_r$. 
At a fixed time moment $t\gtrsim t_r$ one can treat 
$r(k,t)$  as the absolute value of the 
sum of $N$ unit vectors on the plane which have 'stochastic' phases. Accordingly, it is again 
natural to rescale the 'length of the random walk path'  
\begin{equation}\label{st}
r(k,t)=\sqrt{N} \,\,x(k,t), 
\end{equation}
with $x(k,t)\sim 1$ for almost any $t\gtrsim t_r$. 
Substitution of (\ref{st}) into \eqref{rho} leads in the thermodynamic limit $N\to\infty$, 
${\mathfrak N}\to\infty$, 
${\mathfrak N}/N=\bar{n}$ to the Poisson distribution 
\begin{equation}\label{rhoP}
\rho(k,t) =\sum_{J=0}^{\infty}|J\rangle\langle J| \frac{[ \bar{n}\, x^2(k,t) ]^J}{J!}\exp{[-\bar{n}\, x^2(k,t)]}
\end{equation}
for the atom density operators at almost any time moments $t\gg t_r$. On the other hand, the time average 
(\ref{timeav}) of the density operator $\overline{\rho(k,t)}$ 
over the infinite time interval was proved to approach to the canonical 
Gibbs distributions in the thermodynamic limit,
see \eqref{avr}. It follows from \eqref{avr} and \eqref{rhoP}, that the reduced density operator $ \rho(k,t) $
of the $k$-th atom never approach to its time average \eqref{avr} at any time moment. 
Similarly, the quantum averages $\langle A_k\rangle(t)={\rm Tr}^{(s,k)}[A_k\, \rho(k,t)]$ of the local observables $A_k$
relating to the $k$-th atom should also strongly fluctuate in time.\begin{footnote}{
For the 0-th atom energy operator $h_k^{(s)}$, the time fluctuations of its quantum expectation values 
$E(0,t)$ are clearly seen in Figure \ref{R2} at $t\,\Omega^2/\Omega_0\gtrsim 30$.
}
\end{footnote}
This means, in turn, that
the time dispersion 
\begin{equation}\label{dis}
D_{JJ'}(\rho) = \overline{ \rho_J(k,t) \rho_{J'}(k,t) }-\overline{\rho_J(k,t)}
\,\,\overline{\rho_{J'}(k,t)}
\end{equation}
of the density operator  matrix elements $\rho_J(k,t)=\langle J|\rho(k,t)|J\rangle$ 
remains considerable and does not vanish
in the thermodynamic limit. Such type of evolution indicating luck of partial 
thermalization in  model (\ref{HC})    was associated in \cite{rut80-1}
with the strong degeneracy of the 
energy level differences, see equation \eqref{res}. Really, the dispersion \eqref{dis}
can be written as
\begin{eqnarray} \nonumber
&& D_{JJ'}(\rho) =\sum_{ n_1,n_1'  }\sum_{n_2,n_2'} 
\langle J|{\rm Tr}^{(r,k)}\left[|n_1\rangle\langle n_1'\|\right]| J\rangle \,\,
\langle J'|{\rm Tr}^{(r,k)}\left[|n_2'\rangle\langle n_2|\right]| J'\rangle \cdot \\ \nonumber
&&\langle n_1|\psi(0)\rangle \langle \psi(0)|n_1'\rangle
\langle n_2'|\psi(0)\rangle \langle \psi(0)|n_2\rangle \cdot\\
&&\left\{ \overline{\exp[i t(E_{n_2}-E_{n_2'}-E_{n_1}+E_{n_1'})]} -
\overline{\exp[i t(E_{n_2}-E_{n_2'})]} \,\,\,\overline{\exp[i t(E_{n_1'}-E_{n_1})]} \label{dis1}
\right\},
\end{eqnarray}
where $E_n$ are  the energy levels (\ref{ele}) in model (\ref{HC}). 
The last line in \eqref{dis1} has the structure
\begin{eqnarray}\nonumber
&&\left\{ \overline{\exp[i t (E_{n_2}-E_{n_2'}-E_{n_1}+E_{n_1'}) ]} -
\overline{\exp[i t(E_{n_2}-E_{n_2'})]} \,\,\,\overline{\exp[i t(E_{n_1'}-E_{n_1})]}\right\}
=\\
&& \delta_{n_1 n_2} \delta_{ n_1' n_2' }- \delta_{n_1 n_2 n_1' n_2'}  + 
C_{n_1 n_2 n_1' n_2'} ,\label{rt}
\end{eqnarray}
where $\delta_{n_1 n_2}$ is the Kronecker delta, and 
\[
\delta_{n_1 n_2 n_1' n_2'}=
\begin{cases}
1, \quad {\rm if} \quad n_1=n_2= n_1'=n_2, \\
0 \quad{\rm otherwise}.  
\end{cases}
\]
The {\it {resonance term}}  $C_{n_1 n_2 n_1' n_2'}$ in the right-hand side of \eqref{rt}
is nonzero, in particular, at $n_1'-n_1=n_2'-n_2=m$, and 
as well at $n_2-n_1=n_2'-n_1'=m$ due to \eqref{res}. It is precisely this term which is 
responsible for the 
strong time fluctuations of the density operator $\rho(k,t)$ in model (\ref{HC}). 

On the other hand, the resonance term $C_{n_1 n_2 n_1' n_2'}$   vanishes  in (\ref{rt}), 
if a weak anharmonic interaction is applied providing for the shifted energy levels that  
\begin{equation} \label{er}
E_{n_1}-E_{n_1'}\ne E_{n_2}-E_{n_2'} \quad {\rm unless}\quad 
\begin{cases}
{\rm either} \quad n_1=n_1'& {\rm and}\quad n_2=n_2' \\
{\rm or} \quad n_1=n_2& {\rm and}\quad n_1'=n_2'.  
\end{cases}
\end{equation}
In this case, one should expect, that the time dispersion $D_{JJ'}(\rho)$ 
would also vanish in the thermodynamic limit $N\to\infty, {\mathfrak N}/N=\bar{n}$.
For $\bar{n}<1$, this was indeed proved in \cite{rut80-1}.
The crucial role of the 'non-resonance condition' \eqref{er} 
for  ability of an isolated quantum system to thermalize was
first established by von Neumann \cite{Neum29,Neumann}, and confirmed later by 
many authors  \cite{Tas98,Lin09,Leb10a}.
\section{Time dispersion of the quantum expectation value of the atom energy \label{S3}}
It is interesting to illustrate the dramatic effect of the nonlinear interaction 
on the time evolution of the quantum expectation value of the energy $E(k,t)$ of a particular atom,
\begin{equation} \label{en0}
E(k,t)=  \langle \psi(t)| h_k^{(s)} |\psi(t)\rangle,
\end{equation}
where 
\[
h_k^{(s)} =\Omega_0\,b_k^\dagger b_k,
\]
is the Hamiltonian of the $k$-th atom, 
and $|\psi(t)\rangle$ is the quantum state of the chain at time $t$ which  evolves from the initial state (\ref{in}).

In the purely harmonic chain (\ref{HC}), the energy $E(k,t)$ is determined by 
(\ref{en}):
\begin{equation}\label{en1}
E(k,t)= 
{ \mathfrak N}\, \Omega_0 \, \frac{r^2(k,t)}{N^2}.
\end{equation}
Taking into account (\ref{r2}), one obtains from \eqref{en1} for the time averages:
\begin{eqnarray}
&& \lim_{N\to\infty} \overline{E(k,t)} = \begin{cases}
2 \Omega_0 \,\bar{n}, & {\rm if}\quad k=0, \\
2 \Omega_0 \,\bar{n}, & {\rm if} \quad  k=N/2, \; {\rm and } \;  N \; {\rm is\; even},\\
 \Omega_0 \,\bar{n} , & {\rm otherwise}, 
\end{cases}  \label{Eav}\\
&&\lim_{N\to\infty} D( h_k^{(s)} )=\begin{cases}
 4\Omega_0^2\,\bar{n}^2 , & {\rm if}\quad k=0, \\
 4\Omega_0^2\,\bar{n}^2,& {\rm if} \quad  k=N/2, \; {\rm and } \;  N \; {\rm is\; even},\\
 \Omega_0^2\,\bar{n}^2 , & {\rm otherwise}, 
\end{cases}
\end{eqnarray}
where
\begin{equation} \label{dis4}
D( h_k^{(s)} ) \equiv\overline{E(k,t)^2}-\left[\overline{E(k,t)}\right]^2.
\end{equation}
Thus, the time dispersion $D( h_k^{(s)} )$ of the quantum expectation value of the 
atom Hamiltonian remains finite in the
thermodynamic limit 
in the purely harmonic chain. 

On the other hand, in the chain perturbed by the extremely weak anharmonic interaction, the
$k$-th atom energy (\ref{en0}) can be determined as
\begin{equation}\label{en3}
E(k,t)= {\rm Tr}^{(s,k)}[h_k^{(s)} \, \rho(k,t) ],
\end{equation}
and the atom reduced density operator
\begin{eqnarray}\nonumber
\rho(k,t)=\frac{{\mathfrak N}!}{N^{\mathfrak N}}\sum_{J=0}^{\mathfrak N} |J\rangle\langle J|
\sum_{P=J}^{\mathfrak N} \frac{P!^2(-1)^{P-J}}{N^PJ!(P-J)!}\,\, 
\sum_{\sum_\tau l_\tau={\mathfrak N}-P}\frac{1}{l!}\times\\
\sum_{\sum_\tau n_\tau={\mathfrak N}}\frac{\exp\{ i [k(n\tau )-t\, E_n]\}}{(n-l)!}
\sum_{\sum_\tau n'_\tau={\mathfrak N}}\frac{\exp \{i [-k(n'\tau )+t\,E_{n'}]\}}{(n'-l)!},\label{r1a}
\end{eqnarray} 
can be obtained from \eqref{r1} 
by replacement 
\begin{eqnarray*}
&&(\omega n) \to E_n = (\omega n)+\delta E_n ,\\
&&(\omega n') \to E_{n'}= (\omega n')+\delta E_{n'} ,
\end{eqnarray*}
corresponding to the small shift of the energy levels.

If the perturbed  energy levels $E_n$ are not degenerate, 
averaging in  
time  of (\ref{en3}) with the atom density operator 
$\rho(k,t) $ given by (\ref{r1a}) provides
\[
\overline{E(k,t)} = \Omega_0 \,\bar{n},
 \]
instead of (\ref{Eav}). In turn, if the energy levels $E_n$ satisfy the non-resonance 
condition (\ref{er}),
the time dispersion (\ref{dis4}) can be written as
\begin{equation}
D( h_k^{(s)} )= \Omega_0^2\, \sum_{J=0}^\infty \sum_{J'=0}^\infty J\,J' \,D_{JJ'}(\rho),
\end{equation}
where $D_{JJ'}(\rho)$ is given by (\ref{s}). Due to the 
equality
\begin{equation} \label{sJ}
\sum_{J=0}^P J \,\frac{(-1)^{P-J}P!}{J!(P-J)!}= 
\begin{cases}
1, & {\rm if}\quad P=1, \\
0, & {\rm otherwise},  
\end{cases}
\end{equation}
several summations can be explicitly performed in (\ref{sJ}).
The result reads as
\begin{eqnarray}\nonumber
&&{D( h_k^{(s)} )}  = {\Omega_0^2} \frac{{\mathfrak N}!^2}{N^{2{\mathfrak N}+2}}
\sum_{\substack{\sum_\tau n_\tau=\sum_\tau n_\tau'={\mathfrak N}\\
n\ne n'\\
\sum_\tau l_\tau=
\sum_\tau k_\tau={\mathfrak N}-1}} \frac{1}{l!(n-l)!(n'-l)!}\,\,\frac{1}{k!(n-k)!(n'-k)!}=\\
&& {\Omega_0^2}\frac{{\mathfrak N}!^2}{N^{2{\mathfrak N}}}\frac{N(N-1)}{N^2}
\sum_{\sum_\tau l_\tau={\mathfrak N}-1}\prod_{\tau}\frac{1}{l_\tau!^2}=
{\Omega_0^2}\frac{{\mathfrak N}!^2}{N^{2{\mathfrak N}}}\frac{N(N-1)}{N^2}
\oint\frac{dx}{2\pi i\,x^{2{\mathfrak N}-1}}[I_0(2x)]^{N},
\label{sB}
\end{eqnarray}
where integration is performed in the complex 
$x$-plane along the circle centered at origin and going 
in the positive direction, and $I_0(t)$ is the modified Bessel function of the first 
kind,
\[
I_0(t)=\sum_{j=0}^\infty \frac{(t/2)^{2j}}{j!^2}.
\]
\begin{figure}[htb]
\centering
\includegraphics[width=.8\linewidth, angle=00]{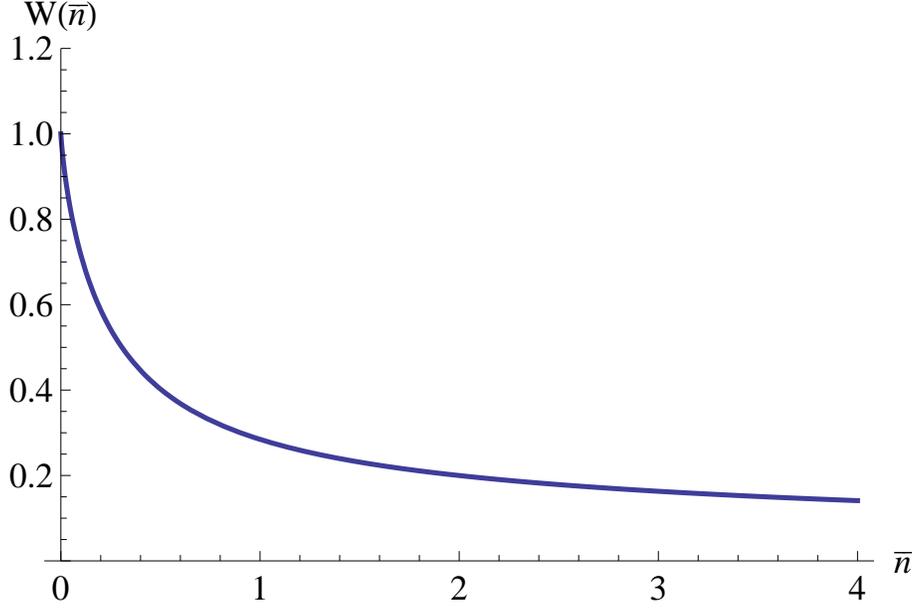}
\caption{\label{WW} Function $W(\bar{n}) $ 
 determined by \eqref{W}.}
\end{figure}
In the thermodynamic limit $N\to\infty$, $\bar{n}={\mathfrak N}/N=Const$,
the integral in the right-hand side of \eqref{sB} is determined by contributions 
of two saddle points at $x=\pm x_0(\bar{n})$,
where $x_0(\bar{n})$ is the positive solution of the equation 
\begin{eqnarray}
\left[\frac{I_1(2 x)}{I_0(2 x)}-\frac{\bar{n}}{x}\right]_{x=x_0(\bar{n})}=0.
\end{eqnarray}
The saddle-point asymptotics of \eqref{sB} in the thermodynamic limit reads as
\begin{equation}\label{spd}
{D( h_k^{(s)} )} =2\,{\Omega_0^2}\,\bar{n}\, x_0(\bar{n})\,\left[
\frac{2\pi N}{f''(x_0)}
\right]^{1/2} [W(\bar{n})]^N \,\left[1+O(N^{-1})
\right],
\end{equation} 
where 
\begin{equation}
f(x)=\log [I_0(2x)]-2\bar{n} \log x,
\end{equation}
and
\begin{equation}\label{W}
W(\bar{n})=\left(\frac{\bar{n}}{e}\right)^{2\bar{n}}\frac{I_0[2 \,x_0(\bar{n})]}
{[x_0(\bar{n})]^{2\bar{n}}}.
\end{equation}

The plot of function $W(\bar{n})$ is shown in Figure \ref{WW}. This function  has 
the following asymptotic behavior
 \begin{eqnarray}
W(\bar{n})=\begin{cases}
\left({\bar{n}}/{e^2}\right)^{\bar{n}}[1+\bar{n}+O(\bar{n}^2)], & \quad \bar{n}\to+0, \\
(4\pi \bar{n})^{-1/2}\,[1+O(\bar{n}^{-1})], & \quad\bar{n}\to+\infty,  
\end{cases}
\end{eqnarray}
and monotonically 
decreases between values $W(0)=1$ and $\lim _{\bar{n}\to\infty}W(\bar{n})=0$ in the half-axis 
$0\le\bar{n}<\infty$. 
Therefore, it follows from 
(\ref{spd}), that  the time dispersion  (\ref{dis4}) of the quantum expectation value 
of the atom energy
vanishes in the thermodynamic limit $N\to\infty$ for all fixed $\bar{n}>0$. 
\section*{Acknowledgements}
I am thankful to P.~Calabrese, J.~Cardy, H.~W. Diehl, D.~Huse, M.~Olshanii and A.~Silva for 
interesting discussions in the 
recent years. I am also grateful to M.~Olshanii and A.~Silva for their advise to publish  the
English translation of \cite{rut80-1}.

\setcounter{footnote}{0}
\newpage
{\LARGE {\bf {Appendix}}}\vspace{1.0cm}
\begin{center}
{\Large{\bf{Relaxation Dynamics of a Quantum Chain \\
of Harmonic Oscillators}}}
\begin{footnote}
{Ukrainian Physical Journal, {\bf 25}, N 7, 1135-1141 (1980).}
\end{footnote}
\\
\vspace{.5cm}
S.~B.~Rutkevich \\
{\small{Kharkov University, Kharkov, Ukraine}}\\
\vspace{.3cm}
{\small {\bf Abstract}}
\end{center}
\par
\begingroup
\leftskip4em
\rightskip\leftskip
{\small
The aim of the present paper is to trace the appearance of some statistical properties
in an exactly solvable dynamic system - a quantum chain with weak harmonic interaction.
The time dependence of the statistical single particle matrices is derived when the system
relaxes from the non-equilibrium pure state. It is shown that the thermodynamic limit
of the matrices in question is the Gibbs distribution. Moreover, the presence of any weak
anharmonic interaction in the system causes each oscillator to have a canonical distribution
almost at any moment.
}
\par
\endgroup
\renewcommand{\theequation}{A.\arabic{equation}}
\setcounter{section}{0}
\setcounter{equation}{0}
\section*{Introduction}
The statistical mechanics, besides the dynamical laws, is based on certain additional
probability hypothesis (the micro-canonical distribution postulate,  the correlation weakness
principle, etc.) which are justified finally by the good agreement of the theory with experiment.
One can guess, that such additional hypothesis are in fact the consequences of the dynamical laws.
This idea was realized in the Boltzmann's approach, for which ergodicity of the Hamiltonian flows plays
an important role. Verification of the latter for real systems is a difficult problem. In \cite{Sinai}
ergodicity of a system of $N$ hard spheres was proved. However, this proof can not be extended to
oscillatory systems. In many papers \cite{Ford,ICh,Z} the ergodicity problem is analyzed analytically and
numerically.

For quantum systems, justification of the micro-canonical distribution is even more problematic,
since the matrix elements of the density operator  calculated in the basis of the stationary
states only change their phases and do not approach to the  equilibrium distribution.
Nevertheless, von Neumann \cite{N} managed to prove some important statements, which could be treated
as quantum basis of the classical statistical mechanics. In the von Neumann's approach, it
is allowed to measure not all dynamical quantities, but only classical (macroscopic) variables
which quantum operators mutually commute.

One can guess, that in a non-isolated system being in  contact with the thermal bath, all
dynamical variables are characterized by statistical properties. In other words, the density
operator of such a system at almost all time moments is the Gibbs distribution.
In the present paper this assumption is verified for the linear chain of harmonic oscillators -
the system of interacting particles, which allows  exact solution of the
Schr{\"o}dinger equation. The dynamics of relaxation in this system is studied as well.
\section{Model}
We consider the periodical chain of harmonic oscillators. Each of them is coupled with
two  neighbours and with its equilibrium point \cite{HT}. The Hamiltonian has the form
\begin{equation}\label{H}
H=\sum_{k=0}^{N-1}\frac{1}{2}\left[p_k^2+U(|q_k-q_{k+1}|)+\Omega_0^2q_k^2\right],
\quad q_N\equiv q_0.
\end{equation}
Here $q_k, \, p_k$ are the coordinate and momentum of the $k$-th oscillator,
$k=0,\ldots,N-1;$ $N$ is the numbers of oscillators, $U(|q_k-q_{k+1}|)/2$ is the
interaction potential of the neighbouring oscillators, $\Omega_0$ is their own frequency,
the Plank constant and the oscillator mass are put  to the unit value, periodical boundary conditions
are chosen. The $k$-th oscillator will be called 'the $k$-th atom'.
Model \eqref{H} has been studied in \cite{MM} in order to calculate correlation functions
and to estimate the Poincare periods in the thermodynamics limit.

The dynamical problem can be solved exactly if $U(|q_k-q_{k+1}|)=\Omega^2(q_k-q_{k+1})^2$. In this case
the normal coordinate $Q_\tau$, $P_\tau$ are introduced as
\begin{eqnarray}
q_k=\sum_\tau e^{ik\tau}Q_\tau/\sqrt{N},\quad
p_k=\sum_\tau e^{-ik\tau}P_\tau/\sqrt{N}, \\
\nonumber
\tau=2\pi l/N, \quad -N/2<l\le N/2,
\end{eqnarray}
with integer $l$. Two sets of secondary quantization operators are related with $q,p,Q,P$:
\begin{eqnarray}\nonumber
b_k=\frac{\Omega_0\,q_k+ip_k}{\sqrt{2\Omega_0}}\quad
b_k\dagger=\frac{\Omega_0\,q_k-ip_k}{\sqrt{2\Omega_0}}
 \\
\label{ab}
a_\tau=\frac{\omega_\tau\,Q_\tau+iP_\tau^\dagger}{\sqrt{2\omega_\tau}}\quad
a_\tau^\dagger=\frac{\omega_\tau\,Q_\tau^\dagger-iP_\tau}{\sqrt{2\omega_\tau}},
\end{eqnarray}
where $\omega_\tau^2=\Omega_0^2+\Omega^2 \, (2\sin\frac{\tau}{2})^2$.

The Hamiltonian can be rewritten in the form
\begin{equation}
H=\sum_\tau(a_\tau^\dagger a_\tau+1/2)\omega_\tau,
\end{equation}
which shows that the normal modes (phonons) do not interact. The
eigenstate of the Hamiltonian reads as
\begin{equation}\label{n}
|n_1,\ldots, n_N\rangle=(n_1!\ldots n_N!)^{-1/2}(a_1^\dagger)^{n_1}\ldots
(a_N^\dagger)^{n_N} |0\rangle,
\end{equation}
where $|0\rangle$ is the ground state, $n_i$ is the number of phonons
with the quasi-momentum $\tau_i$, $\tau_i<\tau_{i+1}$. If $\Omega\ll \Omega_0$,
each atom can be considered as a subsystem which interacts weakly
with surrounding (the thermal bath).
\section{Relaxation of the one-particle density operator}
If $\Omega$ is  small, one can follow the relaxation dynamics of the
single-atom density operator for the system prepared in some realistic
non-equilibrium initial state.

Let us introduce the following notations;
\begin{eqnarray} \nonumber
a^n\equiv \prod_\tau a_\tau^{n_\tau}, \quad
n\equiv \{n_\tau\}, \quad
|n\rangle\equiv |n_1\ldots,n_N\rangle,\quad n!\equiv \prod_\tau n_\tau!,\\
C_n^l\equiv \frac{n!}{l!(n-l)!}, \quad {\rm etc.} 
\end{eqnarray}

If the whole system is described by the density operator
$\rho=\rho_{nn'}|n\rangle\langle n'|$, than the atom with
 the number $k=0$ in the chain has the density operator
${\rm Tr}^{(r)}\rho =\rho_{nn'}{\rm Tr}^{(r)}|n\rangle\langle n'|$, where
\begin{eqnarray*}
{\rm Tr}^{(r)}A=\sum_{m=0}^\infty\sum_{m'=0}^\infty
\sum_{m_1,\ldots ,m_{N-1}=0}^\infty\frac{1}{\sqrt{m_0!m_0'!}m_1!\ldots m_{N-1}!}
(b_0^\dagger)^{m_0}|0\rangle \times
\\
\langle 0|b_0^{m_0}b_1^{m_1}\ldots A\dots (b_1^\dagger)^{m_1}(b_0^\dagger)^{m_0'}
|0\rangle\langle0|b_0^{m_0'}.
\end{eqnarray*}
We have calculated the quantities ${\rm Tr}^{(r)}|n\rangle\langle n'|$
only at $\Omega=0$:
\begin{equation} \label{Tr}
{\rm Tr}^{(r)}|n\rangle\langle n'|=\sum_J |J\rangle\langle J|
\sum_P \frac{P!^2(-1)^{P-J}}{N^P J! (P-J)!}
\sum_{\sum_\tau l_\tau={\mathfrak N}-P}\frac{\sqrt{n!n'!}}{l!(n-l)!(n'-l)!},
\end{equation}
where $|J\rangle=\frac{(b_0^\dagger)^J}{J!}|0\rangle$, ${\mathfrak N}=\sum_\tau n_\tau=
\sum_\tau n_\tau'$. The state $|n\rangle$ has the energy $E_n=E_0+\Omega_0 {\mathfrak N}$,
where $E_0$ is the ground state energy.

Using (\ref{Tr}) it is possible to show that the atom's density operator which corresponds to the
stationary state $|n\rangle$ of the  system, has the form of the Gibbs distribution in
the thermodynamic limit ($N\to\infty$, ${\mathfrak N}/N=\bar{n}={\rm const}$) at least for
$\bar{n}<1$, if the number $\alpha N$ of nonzero integers in the set $n=\{n_\tau\}$ is
macroscopic (i.e. $\alpha\sim 1$):
\begin{equation} \label{Tr1}
{\rm Tr}^{(r)}|n\rangle\langle n|=\frac{1}{1+\bar{n}}\sum_J
\left(\frac{\bar{n}}{1+\bar{n}}\right)^J|J\rangle\langle J|
\left[1+O(N^{-1})\right].
\end{equation}
Perhaps, this result holds to some extent in other dynamical systems.

Let us show, how the evolution of the system from some non-equilibrium
initial state can be described. Let
\begin{equation}
\Omega(t)=
\begin{cases}
  0, \quad  t\le 0, \\
  \Omega, \quad 0<t<T,   \\
0, \quad   t\ge T,
\end{cases}
\end{equation}
and $\Omega$ is so small, that one can neglect the terms of order
$\epsilon$
in
\[
\langle \,n'[\Omega(t')=\Omega]|\,n[\Omega(t)=0]\,\rangle=\delta_{nn'}+\epsilon_{nn'},
\]
i.e. interaction almost does not change the stationary states $|n\rangle$.
At $t\le 0,\, t\ge T$ one can use expression (\ref{Tr}),
and at $0<t<T$ the system relaxes due to the weak interaction.

If at $t=0$ the system is in the stationary state $|n\rangle$, then at time
$T$ its state vector is given by  $|n\rangle \exp[-i(n\omega)T]$ (the terms
of order $\epsilon$ are omitted), where $(\omega n)=\sum_\tau\omega_\tau n_\tau$.

Let us choose the non-equilibrium initial state as
\begin{equation} \label{in}
|\psi(t)|_{t=0}\rangle=\frac{(b_0^\dagger)^{\mathfrak N}}{\sqrt{{\mathfrak N}!}}|0\rangle=
\frac{1}{\sqrt{N^{\mathfrak N}{\mathfrak N}!}}
\sum_{\sum_\tau n_\tau ={\mathfrak N}}\frac{{\mathfrak N}!}{n!}(a^\dagger)^n|0\rangle,
\end{equation}
the 0-th atom is exited to the energy ${\mathfrak N}\Omega_0$, while all the rest atoms
remain unexcited. Then one gets from (\ref{Tr}), (\ref{in})
\begin{eqnarray}\nonumber
\rho(k,t)|_{t=T}=\frac{{\mathfrak N}!}{N^{\mathfrak N}}\sum_J|J\rangle\langle J|
\sum_P\frac{P!^2(-1)^{P-J}}{N^PJ!(P-J)!}\,\, \sum_{\sum_\tau l_\tau={\mathfrak N}-P}\frac{1}{l!}\times\\
\sum_{\sum_\tau n_\tau={\mathfrak N}}\frac{\exp \{i [k(n\tau )-t(n\omega )]\}}{(n-l)!}
\sum_{\sum_\tau n'_\tau={\mathfrak N}}\frac{\exp\{ i [-k(n'\tau )+t(n'\omega )]\}}{(n'-l)!},\label{r1}
\end{eqnarray}
where $\rho(k,t)$ is the density operator of the $k$-th atom ($k=0,1,\ldots,N-1)$ at the
time $t$, and $(n\tau)=\sum_\tau n_\tau \tau$.
Straightforward calculations yield
\begin{equation}\label{rho}
\rho(k,t)=\sum_J|J\rangle\langle J| \frac{{\mathfrak N}!}{J!({\mathfrak N}-J)!}
\left(\frac{r}{N}\right)^{2J}\left[1-\left(\frac{r}{N}\right)^{2}\right]^{{\mathfrak N}-J},
\end{equation}
where
\[
r(k,t)=\left|\sum_\tau \exp i(k\tau-\omega_\tau t)\right|.
\]

Keeping the linear terms in $(\Omega/\Omega_0)^2$ in the expansion of $\omega_\tau$, and replacing the
sum in $\tau$ in $r(q,t)$ by the integral, we come to  the integral representation of the
Bessel function of the $q$-th order:
\begin{equation}
r(k,t)=\left|N J_k[\Omega_0 (\Omega/\Omega_0)^2 t]\right|.
\end{equation}
One can safely replace the sum by the integral at large $N$ and if
$t\ll N/[\Omega_0(\Omega/\Omega_0)^2]$ (the period of the integrand is much large than
$\Delta \tau=2\pi/N$).

One can easily see, that the energy of the $k$-th atom at the time $t$ exceeds
the ground state energy by the quantity
\begin{equation}\label{En}
\Omega_0\langle J\rangle_{\rho(k,t)}=\Omega_0{\mathfrak N}
J_k^2[\Omega_0 (\Omega/\Omega_0)^2 t].
\end{equation}

Perturbation reaches the $k$-th atom after the time of order
$k/[\Omega_0 \,(\Omega/\Omega_0)]^2$, and after that the energy
of this atom oscillates slowly with decreasing amplitude.
When the excitation  energy distributes throughout the whole
chain, one can not any more replace the sum (\ref{rho}) by the integral.
 For the classical harmonic chain, formula (\ref{En})
was obtained in \cite{MM}.
\section{Time average and dispersion of one-particle density operator}
Let us introduce the following notations for the average quantities:
\begin{eqnarray*}
\langle A\rangle_{\rho(t)}={\rm {Tr}} A\rho(t),\\
\bar{f}=\lim_{T\to\infty}\frac{1}{T}\int_0^Tf(t)dt,
\end{eqnarray*}
where $A, \, f$ are an operator a function of time, respectively.

In order to calculate $\langle \bar{A}\rangle_{\rho}$, it is sufficient
to know $\bar{\rho}$. Consider the space of phases with points
$\varphi=(\varphi_1,\ldots,\varphi_N)$, $0\le\varphi_i<2\pi$ (we use the
same notations for indexes as in formula (\ref{n}), and the measure
$d\varphi =\prod_\tau d\varphi_\tau /(2\pi)$.  Let us define the operator-valued
function $\rho(k,\varphi)$ on this space which is obtained from
(\ref{r1}) by the replacement
\[
t(n\omega)\to (n\varphi)=\sum_\tau n_\tau \varphi_\tau,
\quad
t(n'\omega)\to (n'\varphi)=\sum_\tau n_\tau \varphi_\tau.
\]

Due to the equality\begin{footnote} {[S.R. 2012] The
first equality in equation (A.14a) is wrong, since it does not 
take into account that  the energy spectrum $E_n=(\omega n)$ of model 
(\ref{H}) is strongly degenerate. The corrected form of this equation is given by formula \eqref{avv}
in Subsection \ref{ddo}. This correction induces also certain modifications into the calculation 
of $\bar{\rho}$ in this page and into the final result \eqref{r6}. These modifications are described in 
Proposition \ref{av} and equation (\ref{avr}) in Subsection \ref{ddo}. Fortunately, the above
mentioned modifications
are minimal, and equation \eqref{r6} holds for all but one or two atoms in the chain, depending 
on the parity of $N$. On the other hand, application of  a weak anharmonic interaction breaking the energy 
spectrum degeneracy  restores the result \eqref{r6}: the time average of the reduced density 
operators of all atoms in the chain is given by the Gibbs distribution with {\it the same temperature}.

}
\end{footnote}
\[
\overline{\exp[it(\omega(n'-n))]}=\langle\exp[i(\varphi(n'-n))]\rangle_\varphi 
=\delta_{nn'},  \quad\quad \quad\quad\quad\quad \quad\quad\quad\quad \quad\quad
\quad\quad \quad\quad(A.14a)
\] 
one can conclude from (\ref{r1}), that the time average of the operator
$\rho(k,t)$ equals to the phase average of the operator $\rho(k,\varphi)$.
This allows us to write the time average of the operator (\ref{rho}) in the form
\begin{equation}\label{r5}
\bar{\rho}=\sum_J|J\rangle\langle J|\,\, C_{\mathfrak N}^J\,\,
\Big\langle[r(\varphi)/N]^{2 J}\left[1-\left[r(\varphi)/N\right]^2\right]^{{\mathfrak N}-J}
\Big\rangle_\varphi,
\end{equation}
where
\[
r(\varphi)=\bigg|\sum_\tau \exp(i\varphi_\tau)\bigg|.
\]
The problem is reduced to the random walk problem, which was studied in \cite{Cha}.
Using formulas (51)-(53),
(103), (104) of reference \cite{Cha} we find
\begin{eqnarray} \nonumber
p_N(r)\equiv \langle \delta \left(r-r(\varphi)\right)\rangle_\varphi
=r \int_0^\infty d\rho \, \rho J_0^N(\rho)\,J_0(r\,\rho), \\
p_N(r)\to \frac{2r}{N} e^{-r^2/N}, \quad {\rm at }\quad N\to\infty. \label{A16}
\end{eqnarray}
Using the above large-$N$ asymptotics for $p_N(r)$ and for the asymptotic formula
\[
\left[1-(r/N)^2\right]^{{\mathfrak N}-J}\to \exp[-r^2 {\mathfrak N}/N^2]
, \quad {\rm at }\quad N\to\infty,
\]
we get
\begin{equation}\label{r6}
\lim_{N\to\infty}\bar{\rho}_N=\sum_J\frac{1}{1+\bar{n}}
\left(\frac{\bar{n}}{1+\bar{n}}\right)^J
|J\rangle\langle J| .
\end{equation}
The canonical distribution is obtained for all values of $\bar{n}$.

It is clear from (\ref{rho}) that the time fluctuations of the density operator do
not vanish in the thermodynamic limit. This behavior is caused by the degeneracy
of the system
 in the differences of the energy levels of the stationary states:
\begin{equation}
(\omega n)-(\omega n')=\big(\omega(n+m)\big)-\big(\omega(n'+m)\big). \label{res}
\end{equation}

If a weak enough  anharmonic interaction is applied
\begin{itemize}
\item[(a)]\label{a}
this will not change substantially $\bar{\rho}$,
\item[(b)]\label{be}
at least at $\bar{n}<1$, the dispersion
\[
D_{JJ'}(\rho)=\overline{(\rho_J-\bar{\rho}_J)^*(\rho_{J'}-\bar{\rho}_{J'})}
\]
vanishes in the thermodynamic limit.
\end{itemize}
A weak anharmonic interaction can be described as a perturbation
leading to a small shift of each energy level
\[
E_n=(\omega n)+\delta E_n.
\]
Therefore, in equation (\ref{r1}), we should replace now $(n\omega)$
by $E_n$, and $(n'\omega)$
by $E_{n'}$, and the statement (a) becomes evident.

Let us prove (b). One obtains from (\ref{rho})
\begin{eqnarray}\nonumber
D_{JJ'}(\rho)= \frac{{\mathfrak N}!^2}{N^{2{\mathfrak N}}}
\sum_{PQ}\frac{P!^2(-1)^{P-J}}{N^PJ!(P-J)!}
\frac{Q!^2(-1)^{Q-J'}}{N^Q\,J'!(Q-J')!}\times \\
\sum_{\substack{\sum_\tau n_\tau=\sum_\tau n_\tau'={\mathfrak N}\\
n\ne n'\\
\sum_\tau l_\tau={\mathfrak N}-P,
\sum_\tau k_\tau={\mathfrak N}-Q}} \frac{1}{l!(n-l)!(n'-l)!}\,\,\frac{1}{k!(n-k)!(n'-k)!}.
\label{s}
\end{eqnarray}
In the sum (\ref{s}), the following inequality holds
$\|n-n'  \|\le 2 \min (P,Q), $ where $\|n\|=\sum_\tau |n_\tau|$.
Since
\begin{eqnarray}\nonumber
\frac{{\mathfrak N}!^2}{N^{2{\mathfrak N}}}
\sum_{\substack{\sum_\tau n_\tau=\sum_\tau n_\tau'={\mathfrak N}\\
n\ne n'\\
\sum_\tau l_\tau={\mathfrak N}-P,
\sum_\tau k_\tau={\mathfrak N}-Q}} \frac{1}{l!(n-l)!(n'-l)!}\,\,\frac{1}{k!(n-k)!(n'-k)!}\le \\
C_{{\mathfrak N}}^PC_{{\mathfrak N}}^Q
\frac{{\mathfrak N}!^2}{N^{2{\mathfrak N}}}
\sum_{\substack{\sum_\tau n_\tau=\sum_\tau n_\tau'={\mathfrak N}\\
0<\|n-n'\|\le 2 \min (P,Q)}} \frac{1}{n!\,n'!}\le\frac{{\mathfrak N}^P {\mathfrak N}^Q }{P! Q!},
\end{eqnarray}
the series in (\ref{s}) uniformly converges at $\bar{n}<1$. Therefore, one can treat $P$ and
$Q$ as finite numbers. Let $P<Q$. The number of terms in the
sum
\[
\sum_{\substack{\sum_\tau n'={\mathfrak N}\\
0<\|n-n'\|\le2 P}}\frac{1}{n'!}
\]
is smaller than
\[
\left[\frac{(N+P-1)!}{P!(N-1)!}\right]^2,
\]
therefore
\begin{eqnarray}
\frac{{\mathfrak N}!^2}{N^{2{\mathfrak N}}}\sum_{\sum_\tau n_\tau ={\mathfrak N}}\frac{1}{n!}
\sum_{\substack{\sum_\tau n'_\tau ={\mathfrak N}\\
0<\|n-n'\|\le 2P}}\frac{1}{n'!}\le
\frac{{\mathfrak N}!^2}{N^{2{\mathfrak N}}}\sum_{\sum_\tau n_\tau ={\mathfrak N}}\frac{1}{n!}
\left[\frac{(N+P-1)!}{P!(N-1)!}\right]^2\le
\frac{\bar{n}^{\mathfrak N}N^{2P}}{P!^2},
\end{eqnarray}
which proves statement ({b}). It is quite possible that (\ref{Tr1}) and  (b) hold also
for $\bar{n}\ge 1$.

In conclusion let us note that the considered model can be useful for many other applications.
Since in the case of the harmonic interaction, its dynamics is integrable, the model
can be used to verify different approximate methods of studying relaxation processes. In the case of the
anharmonic interaction, one can study analytically or numerically the relaxation dynamics in
a nonlinear system. The model could be useful also for analyzing the arising statistical properties
in a non-macroscopic systems, like three-atom molecules.

The author is grateful to S.~V.~Peletminski for the discussion of the results of this work.

\end{document}